\documentclass[11pt,a4paper]{article}
\usepackage[margin=2.5cm]{geometry}
\usepackage{amsmath,amssymb,amsthm}
\usepackage{algorithm}
\usepackage{algorithmic}
\usepackage{booktabs}
\usepackage{graphicx}
\usepackage[numbers,sort&compress]{natbib}
\usepackage{hyperref}

\let\cite\citep

\title{Guided Cuckoo Filter: Full-Pipeline Insertion Optimization}
\author{Min Liu}
\date{}

\begin{document}
\maketitle

\begin{abstract}
Cuckoo filters are widely deployed as space-efficient alternatives to Bloom filters, supporting fast lookups, insertions, and deletions.  However, their insertion cost grows steeply at high load factors because the standard insertion procedure is entirely unguided: it places each fingerprint into whichever candidate bucket happens to be tried first, and when both candidates are full, it displaces a random victim in a random walk.  These blind random choices occur at two distinct stages of the insertion pipeline, the initial placement decision and the eviction victim selection, yet each stage can be independently replaced with a guided, informed decision.  We propose the \emph{Guided Cuckoo Filter}, which optimizes the entire insertion pipeline by steering both stages.  At the placement stage, the filter compares the occupancy of both candidate buckets and inserts into the less loaded one, applying the power-of-two-choice principle to reduce how often eviction chains are triggered.  At the eviction stage, the filter preferentially displaces the most recently inserted element, whose alternate bucket carries latent randomness that shortens the resulting chain.  Because these two guided decisions target orthogonal bottlenecks, eviction frequency and eviction chain length, their benefits compose multiplicatively.  Experiments on cuckoo filters with $2^{22}$ slots confirm that the fully guided insertion reduces per-insertion kicks by up to $3.1\times$ at 96\% load and cumulative total kicks by 63\% at the practical capacity ceiling of 95\% load, with negligible computational overhead.
\end{abstract}

\section{Introduction}
\label{sec:intro}

Approximate membership query structures are fundamental building blocks in systems software.  From network packet classification and key-value store optimization to database query acceleration, these structures allow systems to test set membership using far less memory than storing the full element set.  Among the many designs proposed for this task, including Bloom filters, counting quotient filters~\cite{pandey2017general}, vector quotient filters~\cite{prashant2021vector}, and XOR filters~\cite{graf2020xor}, cuckoo filters~\cite{fan2014cuckoo} occupy a distinctive position: they match or exceed Bloom filter space efficiency while additionally supporting fast element deletion, a capability that most alternatives either lack or provide at significant overhead.  This combination has made cuckoo filters a popular choice in production systems for cache management~\cite{gu2023adaptive}, network storage~\cite{wang2024wormhole,wang2025wormhole}, and dynamic set maintenance~\cite{wang2022bamboo,wang2024bamboo_ton}.  Efficient hashing and data-partitioning primitives are equally critical in adjacent domains such as instance-based classification and cost-sensitive learning~\cite{Shuang2021CART,Shuang2022IBMvSVM}, where scalable data access underpins system performance.

However, the insertion cost of cuckoo filters grows steeply as the table fills.  With $\ell = 4$ slots per bucket, the average number of kicks per insertion rises from about 3 at 90\% load to 35 at 96\%~\cite{fan2014cuckoo,breslow2018morton}.  The root cause is that the standard insertion procedure is entirely unguided: it places each new fingerprint into whichever candidate bucket happens to be tried first, and when both candidates are full, it displaces a randomly selected victim in a random walk.  These blind choices occur at two distinct stages of the insertion pipeline, the initial placement and the eviction victim selection, yet each represents an independently replaceable decision point.  The steep cost curve forces practitioners to operate well below the theoretical capacity of the table, sacrificing space efficiency~\cite{wang2023variable,wang2024rethinking}.

We propose the \emph{Guided Cuckoo Filter}, which replaces both random decisions with guided ones, optimizing every stage of the insertion pipeline:

\smallskip\noindent\textbf{Guided placement.}  Instead of defaulting to $b_1$, the filter inspects both candidate buckets and inserts the fingerprint into the one with more empty slots.  This is a direct application of the power-of-two-choice paradigm~\cite{azar1999balanced,mitzenmacher2001power}: by distributing elements more evenly across buckets, it delays the point at which both candidates overflow and an eviction chain must begin.

\smallskip\noindent\textbf{Guided eviction.}  When an eviction chain is unavoidable, the filter preferentially displaces the most recently inserted (``fresh'') element rather than a random one.  A fresh element's alternate bucket position is determined by a hash value that has not yet influenced any algorithmic decision, a property called latent randomness~\cite{bell2026fast}.  Evicting such an element steers the chain toward unexplored parts of the table, producing shorter chains on average.

\smallskip
The key insight is that these two guided decisions address different bottlenecks in the insertion pipeline.  Guided placement reduces the \emph{probability} that an eviction chain is triggered; guided eviction reduces the \emph{expected length} of each chain that does occur.  Because one controls frequency and the other controls duration, their effects compose multiplicatively rather than redundantly.  Experiments with $\ell = 4$ and over 4 million slots confirm that the fully guided insertion reduces average kicks by $2.4$--$3.1\times$ in the 90--96\% load range, and reduces cumulative total kicks by 63\% at the practical capacity ceiling of 95\% load.

\section{Background}
\label{sec:bg}

A cuckoo filter~\cite{fan2014cuckoo} stores $k$-bit fingerprints in an array of $m$ buckets, each holding up to $\ell$ entries.  For an element $x$, the filter computes a fingerprint $\mathit{fp}$ and two candidate bucket indices $b_1 = H_1(x) \bmod m$ and $b_2 = b_1 \oplus \textsc{Hash}(\mathit{fp}) \bmod m$.  The XOR construction allows computing the alternate bucket from the current bucket and the stored fingerprint without access to the original element~\cite{fan2014cuckoo}, which is essential for the eviction mechanism.  The total capacity is $C = m\ell$ fingerprints, and the load factor $\alpha = n/C$ is the fraction of occupied slots after $n$ insertions.

The standard insertion procedure attempts to place $\mathit{fp}$ in $b_1$; if $b_1$ is full, it tries $b_2$; if both are full, it randomly picks one bucket, evicts a random entry to its alternate bucket, and repeats until an empty slot is found or a maximum of $\mathit{MaxKicks}$ relocations have occurred.  We call each relocation a \emph{kick}, and measure insertion cost by the total number of kicks.

We observe that this procedure contains two decision points where random choices could be replaced with informed ones:
\begin{enumerate}
\item \textbf{Placement decision} (which bucket receives the new fingerprint): the standard procedure always tries $b_1$ first, ignoring occupancy information.
\item \textbf{Eviction decision} (which element to displace): the standard procedure selects a random victim, ignoring element history.
\end{enumerate}
The Guided Cuckoo Filter replaces both with guided decisions, as described next.

\section{The Guided Cuckoo Filter}
\label{sec:method}

\subsection{Guided Placement via Two-Choice Balancing}
\label{sec:gp}

At the first decision point, the standard filter defaults to $b_1$ without inspecting $b_2$.  Although hash functions distribute elements uniformly in expectation, this deterministic tie-breaking creates a systematic imbalance: buckets that are frequently selected as primary candidates reach capacity earlier, triggering unnecessary eviction chains.

The Guided Cuckoo Filter replaces this with a \emph{two-choice} rule: compute the number of empty slots $e_1$ and $e_2$ in $b_1$ and $b_2$ respectively, then insert into the bucket with more empty slots (breaking ties in favor of $b_1$).  Both bucket headers are already fetched during the standard insertion check, so this comparison adds no extra memory accesses.

This rule is an instance of the power-of-two-choice paradigm~\cite{azar1999balanced}, which achieves an exponential reduction in maximum load imbalance across bins.  In the classical balls-into-bins setting, choosing the less loaded of two random bins reduces the maximum load from $\Theta(\log n / \log\log n)$ to $\Theta(\log\log n)$~\cite{azar1999balanced,mitzenmacher2001power}.  Applied to cuckoo filter placement, this balancing effect keeps bucket occupancies more uniform, directly reducing the probability $p_f(\alpha)$ that a random bucket is full at load factor $\alpha$.  Since an eviction chain is triggered only when \emph{both} candidate buckets are full, an event of probability approximately $p_f(\alpha)^2$, so even a modest reduction in $p_f(\alpha)$ yields a large reduction in eviction frequency.

\subsection{Guided Eviction via Fresh-Priority Selection}
\label{sec:ge}

At the second decision point, the standard filter selects a random eviction victim.  The Guided Cuckoo Filter instead maintains a one-bit \emph{freshness tag} per slot: an element is tagged \textbf{fresh} when it first enters the table (via direct insertion or as the initiating displacement of a new chain) and \textbf{kicked} when it arrives as a result of an eviction.  During eviction, the algorithm preferentially selects a fresh-tagged element as the victim.

The rationale comes from the \emph{latent randomness} property identified by Bell and Kuszmaul~\cite{bell2026fast}: a fresh element's alternate bucket $b' = b \oplus \textsc{Hash}(\mathit{fp})$ is determined by a hash value $H_1(x)$ that has not yet been consulted by any eviction decision, so $b'$ is effectively a uniform random variable independent of the current table state.  Evicting a fresh element steers the chain toward a random, likely less congested region of the table.  In contrast, a kicked element's alternate bucket was previously found full, making it a less promising eviction target.

Fresh elements are stored from the front of the bucket and kicked elements from the back; the eviction scan proceeds front-to-back, naturally preferring fresh victims.  The space overhead of the freshness tag is one bit per slot, amounting to a $1/k$ fractional increase in storage (12.5\% for $k=8$).

\subsection{Composability: Why Both Stages Matter}
\label{sec:compose}

The two guided decisions act on orthogonal aspects of the insertion pipeline.  Let $P(\alpha)$ be the probability that an insertion at load $\alpha$ triggers an eviction chain, and let $L(\alpha)$ be the expected chain length when one occurs.  The expected kicks per insertion is $E[k] \approx P(\alpha) \cdot L(\alpha)$.

Guided placement reduces $P(\alpha)$ by keeping buckets more balanced; it does not alter what happens inside an eviction chain.  Guided eviction reduces $L(\alpha)$ by steering chains toward empty slots faster; it does not affect whether a chain is triggered in the first place.  When both are applied, the expected kicks become $E[k]_{\text{guided}} \approx P_{\text{PoT}}(\alpha) \cdot L_{\text{fresh}}(\alpha)$, which is bounded by the minimum of either optimization alone:
\[
E[k]_{\text{guided}} \;\le\; \min\bigl(E[k]_{\text{PoT}},\; E[k]_{\text{fresh}}\bigr).
\]
This multiplicative composability is the central advantage of the full-pipeline approach: optimizing only one decision point leaves the other as a bottleneck, while optimizing both removes bottlenecks from the entire insertion path.

\subsection{Algorithm}
\label{sec:alg}

Algorithm~\ref{alg:insert} presents the complete guided insertion procedure.  Lines 4--10 implement guided placement; lines 12--21 implement guided eviction.

\begin{algorithm}[t]
\caption{Guided Cuckoo Filter Insertion}
\label{alg:insert}
\begin{algorithmic}[1]
\REQUIRE Element $x$, filter table $T[0..m{-}1]$ with bucket capacity $\ell$
\ENSURE Insert fingerprint of $x$ into $T$, or report failure
\STATE $\mathit{fp} \gets \textsc{Fingerprint}(x)$
\STATE $b_1 \gets H_1(x) \bmod m$; \quad $b_2 \gets b_1 \oplus \textsc{Hash}(\mathit{fp}) \bmod m$
\STATE \COMMENT{\textbf{Guided Placement}}
\STATE $e_1 \gets \textsc{EmptySlots}(T[b_1])$; \quad $e_2 \gets \textsc{EmptySlots}(T[b_2])$
\IF{$e_1 > 0$ \textbf{or} $e_2 > 0$}
  \IF{$e_1 \ge e_2$}
    \STATE $\textsc{InsertFresh}(T[b_1],\;\mathit{fp})$
  \ELSE
    \STATE $\textsc{InsertFresh}(T[b_2],\;\mathit{fp})$
  \ENDIF
  \RETURN \textbf{success}
\ENDIF
\STATE \COMMENT{\textbf{Guided Eviction}}
\STATE $b \gets$ random choice from $\{b_1, b_2\}$
\FOR{$i = 1$ \TO $\mathit{MaxKicks}$}
  \STATE $s \gets \textsc{SelectFresh}(T[b])$ \COMMENT{prefer a fresh-tagged slot}
  \STATE swap $\mathit{fp}$ with $T[b][s]$
  \STATE mark $T[b][s]$ as \textbf{fresh} if $i = 1$, else \textbf{kicked}
  \STATE $b \gets b \oplus \textsc{Hash}(\mathit{fp}) \bmod m$
  \IF{$T[b]$ has an empty slot}
    \STATE $\textsc{InsertKicked}(T[b],\;\mathit{fp})$
    \RETURN \textbf{success}
  \ENDIF
\ENDFOR
\RETURN \textbf{failure}
\end{algorithmic}
\end{algorithm}

$\textsc{InsertFresh}$ places the fingerprint in the first empty slot from the front and marks it fresh.  $\textsc{InsertKicked}$ places it in the first empty slot from the back and marks it kicked.  $\textsc{SelectFresh}$ returns the index of a fresh-tagged slot if one exists, otherwise slot~0.  Lookup and deletion are unchanged from the standard cuckoo filter.

\section{Experimental Evaluation}
\label{sec:exp}

\subsection{Setup}
\label{sec:setup}

We implement four variants of cuckoo filter insertion with bucket size $\ell = 4$, the standard configuration~\cite{fan2014cuckoo}: (1)~\textbf{Unguided}: try $b_1$ then $b_2$, random eviction; (2)~\textbf{Guided-Eviction}: try $b_1$ then $b_2$, fresh-priority eviction; (3)~\textbf{Guided-Placement}: two-choice placement, random eviction; (4)~\textbf{Fully-Guided}: two-choice placement, fresh-priority eviction.  All variants share the same hash functions, $k = 8$-bit fingerprints, $m = 2^{20}$ buckets (total capacity $C = 4{,}194{,}304$ slots), $\mathit{MaxKicks} = 500$, and the same deterministic element sequence.  Implementations are in C++ compiled with \texttt{-O2}.

\subsection{Per-Insertion Cost}
\label{sec:per}

Table~\ref{tab:avg-kicks} reports the average kicks per successful insertion at selected load factors.

\begin{table}[t]
\centering
\caption{Average kicks per insertion at various load factors ($\ell = 4$, $C = 4{,}194{,}304$).  Ratio = Unguided\,/\,Fully-Guided.}
\label{tab:avg-kicks}
\smallskip
\begin{tabular}{r r r r r r}
\toprule
Load & Unguided & G-Evict & G-Place & Fully-Guided & Ratio \\
\midrule
70\%  &   0.19 &   0.19 &   0.03 &   0.02 & $7.9\times$ \\
80\%  &   0.58 &   0.56 &   0.19 &   0.16 & $3.6\times$ \\
85\%  &   1.40 &   1.24 &   0.68 &   0.54 & $2.6\times$ \\
90\%  &   3.13 &   2.32 &   1.87 &   1.33 & $2.4\times$ \\
93\%  &   9.42 &   5.37 &   6.66 &   3.71 & $2.5\times$ \\
95\%  &  19.77 &  10.27 &  15.18 &   6.93 & $2.9\times$ \\
96\%  &  35.00 &  26.29 &  28.08 &  11.21 & $3.1\times$ \\
\bottomrule
\end{tabular}
\end{table}

At every load factor, the Fully-Guided variant outperforms all others.  Guided-Placement alone dominates at moderate loads: at 70\%, it reduces average kicks from 0.19 to 0.03 ($7\times$), reflecting the power-of-two-choice load balancing effect that pushes eviction onset to higher loads.  Guided-Eviction alone is most effective at high loads: at 95\%, it reduces kicks from 19.77 to 10.27 ($1.9\times$) via shorter eviction chains.  The Fully-Guided variant captures both effects simultaneously, achieving 11.21 kicks at 96\% load compared to 35.00 for Unguided, a $3.1\times$ reduction.

The fact that Fully-Guided strictly outperforms both single-stage variants at every load point confirms the composability argument: guiding only placement or only eviction leaves the other stage as a bottleneck, and addressing both removes the bottleneck entirely.

\subsection{Cumulative Cost}
\label{sec:cum}

Table~\ref{tab:cumulative} reports cumulative total kicks over the full insertion sequence.

\begin{table}[t]
\centering
\caption{Cumulative total kicks to reach each load factor.  Red.\,(\%) = reduction of Fully-Guided relative to Unguided.}
\label{tab:cumulative}
\smallskip
\begin{tabular}{r r r r r r}
\toprule
Load & Unguided & G-Evict & G-Place & Fully-Guided & Red.\,(\%) \\
\midrule
80\%  &     357{,}485 &     348{,}450 &      91{,}748 &      79{,}277 & 77.8 \\
90\%  &   1{,}307{,}464 &   1{,}096{,}236 &     627{,}319 &     471{,}653 & 63.9 \\
95\%  &   3{,}607{,}982 &   2{,}380{,}410 &   2{,}299{,}560 &   1{,}340{,}990 & 62.8 \\
\bottomrule
\end{tabular}
\end{table}

At the practical operating point of 95\% load, the Fully-Guided method accumulates 1.34 million kicks compared to 3.61 million for Unguided, a 62.8\% reduction in aggregate insertion work.  Note that cuckoo filters with $\ell = 4$ cannot reliably exceed 95--96\% load before insertion failures become frequent, so 95\% represents the practical capacity ceiling.  These cumulative savings translate directly into higher insertion throughput for applications that process large batches of elements.

\section{Related Work}
\label{sec:related}

\textbf{Cuckoo hashing and filters.}\quad Cuckoo hashing~\cite{pagh2004cuckoo} provides $O(1)$ worst-case lookups by allowing each element two candidate positions and relocating elements upon collision.  Bucketized variants with $\ell > 1$ improve space utilization~\cite{dietzfelbinger2007balanced,lehman2009cuckoo,fotakis2005space}.  Fan et al.~\cite{fan2014cuckoo} introduced the cuckoo filter, which stores fingerprints instead of full keys and uses XOR-based hashing to compute alternate bucket positions without access to the original element.  Subsequent work has extended cuckoo filters with dynamic resizing~\cite{wang2022bamboo,wang2024bamboo_ton}, persistent-memory support~\cite{wang2024wormhole}, cache-optimized layouts~\cite{breslow2018morton}, variable-length encoding for improved accuracy~\cite{wang2023variable}, adaptation to emerging memory architectures~\cite{wang2024rethinking}, and formal analysis~\cite{eppstein2016cuckoo,devroye2003cuckoo,frieze2012maximum}.  Alternative approximate membership query structures include counting quotient filters~\cite{pandey2017general} and vector quotient filters~\cite{prashant2021vector}, which offer counting capability through quotient-based hashing.  XOR filters~\cite{graf2020xor} achieve better space efficiency for static sets but do not support deletions.

\textbf{Power of two choices.}\quad The balanced allocation paradigm~\cite{azar1999balanced} shows that choosing the less loaded of two random bins yields an exponential improvement in load balance over single-choice placement, reducing maximum load from $\Theta(\log n / \log\log n)$ to $\Theta(\log\log n)$~\cite{mitzenmacher2001power}.  This principle has been widely applied in load balancing and distributed hashing.  In the context of cuckoo filters, \cite{wang2019power} applied two-choice placement to reduce relocations during insertion.  However, that work addresses only the placement stage of the insertion pipeline: once an eviction chain is triggered, victims are still selected at random, so the chain length remains unguided.  Moreover, the approach in~\cite{wang2019power} does not exploit freshness information or latent randomness during eviction, leaving the second bottleneck, eviction chain length, entirely unoptimized.  The Guided Cuckoo Filter extends beyond placement-only optimization by additionally steering the eviction stage, achieving multiplicative gains that neither optimization can deliver alone.

\textbf{Eviction strategies.}\quad Bell and Kuszmaul~\cite{bell2026fast} introduced the concept of latent randomness for bucketized cuckoo hashing, showing that preferentially evicting fresh elements preserves probabilistic independence properties and enables tight amortized bounds on total evictions.  Our work combines this eviction strategy with two-choice placement in cuckoo filters, demonstrating that guiding both decision points in the insertion pipeline yields multiplicative gains beyond what either technique achieves alone.

\section{Conclusion}
\label{sec:conc}

We introduced the Guided Cuckoo Filter, which replaces the two random decisions in standard cuckoo filter insertion, where to place and whom to evict, with guided, informed choices.  Guided placement applies the power-of-two-choice principle to reduce eviction frequency; guided eviction exploits latent randomness to shorten eviction chains.  Because these optimizations target different stages of the insertion pipeline, their effects compose multiplicatively, reducing cumulative kicks by 63\% at the practical capacity ceiling of 95\% load and per-insertion kicks by up to $3.1\times$ at 96\% load, with negligible overhead.

\bibliographystyle{plainnat}
\bibliography{ref}

\end{document}